\documentclass[12pt,preprint]{aastex}
\usepackage{natbib}

\shorttitle{Evolution of Bipolar Outflows from YSOs}
\shortauthors{Seale et al.}

\begin{document}

\title{Morphological Evolution of Bipolar Outflows from YSOs}

\author{Jonathan P. Seale and Leslie W. Looney}
\affil{Astronomy Department, University of Illinois,
    Urbana, IL 61801}
\affil{Department of Astronomy, MC-221 \\ 1002 W. Green Street \\ Urbana, IL 61801}

\begin{abstract}

We present Spitzer IRAC images that indicate the presence of cavities cut into the dense outer envelope surrounding very young pre-main sequence stars. These young stellar objects (YSOs) characterized by an outflow represent the earliest stages of star formation. Mid-infrared photons thermally created by the central protostar/disk are scattered by dust particles within the outflow cavity itself into the line of sight. We observed this scattered light from 27 nearby, cavity-resolved YSOs, and quantified the shape of the outflow cavities. Using the grid models of \citet{rob06}, we matched model spectral energy distributions (SEDs) to the observed SEDs of the 27 cataloged YSOs using photometry from IRAC, MIPS, and IRAS. This allows for the estimation of geometric and physical properties such as inclination angle, cavity density, and accretion rate. By using the relative parameter estimates determined by the models, we are able to deduce an evolutionary picture for outflows. Our work supports the concept that cavities widen with time, beginning as a thin jet-like outflow that widens to reveal the central protostar and disk until the protostellar envelope is completely dispersed by outflow and accretion.

\end{abstract}


\section{Introduction}
In even their earliest stages of formation, young stellar objects (YSOs) are known to produce outflows that perturb their surrounding parental gas cloud \citep[e.g.][]{shuetal87}. Due to extreme obscuration, for the youngest, most embedded sources, these outflows may be one of the best means of probing the physical conditions of star formation. An outflow can have a profound effect on its parent cloud as jets and prestellar winds carve cavities and inject energy and momentum into the surrounding medium \citep[e.g.][]{shu00,bal07}. The presence of an outflow can be influential enough to disrupt neighboring cloud cores through turbulent motions \citep{bal96}. With protostellar winds that originate close to the surface of the forming star \citep[e.g.][]{shu95,sha07,pud07}, outflows may not only have an effect on the surrounding material but on the final properties of the newborn star as well. Outflows clear out gas and dust from the protostellar surroundings, resulting in the termination of the infall stage \citep{vel98} and affecting the star formation efficiency of the cloud \citep{mat00}. 

A general evolutionary picture of star formation has been formed over the years to be consistent with observations of young stars \citep{lad84,ada86,ada87}. The current picture of star formation begins with the densest parts of a molecular cloud gravitationally and quasi-statically contracting to form a dense prestellar core. Collapse can only take place once gravity has overtaken internal turbulent and/or magnetic support within the collapsing cloud region. As the core continues to gravitationally collapse, a central protostar forms surrounded by an infalling envelope and self-supported protostellar disk. Initially, the central object is completely shielded by the dense envelope, which functions as the primary mass reservoir for the accreting star. Over time, the surrounding envelope loses mass to both infall onto the central object/disk and outflow into the surrounding interstellar medium. Eventually, the central young star and prestellar disk are revealed once the envelope is completely dispersed by both outflow and accretion.

Because this evolutionary sequence is fluid, identifying specific evolutionary stages can be difficult. The most popular categorization system is a class scheme (Class 0, I, II, III) \citep{lad87,and93}. Class 0, the earliest evolutionary stage, is characterized by the densest infalling envelope and an outflow. This outflow is often times jet-like or collimated in nature $-$ having a very small opening angle. More evolved sources (Class I) have larger outflows possibly consisting of a jet and a broader protostellar wind outflow component \citep{arc06,shu00}. The surrounding envelope's density begins to decrease as mass is lost to the central protostar or is swept away by the outflow. By Class II, the protostellar disk has grown in mass, and the outflow cavity has greatly widened and may no longer have a definite structure \citep{bac96}. The most evolved stage of YSO, a Class III object consists solely of a T Tauri, pre-main sequence, star and either an optically thin disk or no disk at all. Some Class III objects may still possess an optical bipolar jet \citep[e.g.][]{bal07}.

\subsection{YSO Outflows}

The presence of an outflow is a fundamental property of young stellar objects; some of the earliest observations revealed that outflows are common around young forming stars \citep[e.g.][]{bal83,edw82,edw83,edw84}. YSOs develop outflows during the earliest stages of evolution; indeed, Class 0 sources are characterized by outflows \citep{and93}. The cavities carved-out by outflows are bipolar \citep{shu87}, signaling the presence of bipolar winds or jets. Bipolar jets or collimated outflows are a ubiquitous phenomenon of astronomical objects involving accretion, rotation, and magnetic fields \citep[e.g.][]{bal07}. Initially during the infall phase, the inflowing material acts to suppress the stellar outflow. Because the total column is the most depleted at the poles \citep{ter84} and the envelope is rotationally flattened, breakout will first occur at the poles \citep{shu87}. Combined with the presence of a thick protostellar disk, these physical characteristics lead to outflows with a symmetry about the poles.

Although never conclusively demonstrated with a very large sample of sources, it is expected that the cavity opening angles must widen over time \citep{shu87,arc06,tob07}. A forming star's evolution, which begins with a heavily embedded protostar and an outflow, must ultimately result in the presence of a protostar and disk without the embedding envelope. Today there is no singular generally accepted explanation for outflow cavity widening. It may be that jet axis precession could produce wider cavities over time \citep{mas93}. However, despite ample evidence of axis wandering \citep{rei97,ter99}, in most cases, the angle of the precession is too small to account for the full extent of the cavity widening \citep{yu99,rei00,arc04}. It has also been suggested that the outflow cavity is formed by both a collimated wind or jet and a broader wide-angle protostellar wind \citep{kwa95,kon00,shu00,mat03}. The X-wind model predicts a high density protostellar wind along the axis, with a decreasing density with increasing angle from the axis \citep{sha98}. According to this model, in very early stages (young Class 0 sources), only the dense, high-velocity, collimated component of the outflow can puncture the surrounding high density envelope \citep{wil03}. Over time, the envelope loses mass, allowing the weaker, less dense, and lower velocity broad-angle component to carve a wider cavity into the envelope. Combined with the entraining of the cavity's edge material out of the envelope, this two-component wind can lead to the widening of the outflow. Additionally, since the accretion rate lessens over time, the decrease in infall eventually allows the outflow to dominate over the infall. This could allow the wind to carve out wider cavities as the system evolves \citep{del00}.

\subsection{Procedure}

YSOs emit most of their radiation at wavelengths longward of the near-infrared \citep[e.g.][]{ada87}. The primary radiation source is thermal in nature, emanating from the envelope, circumstellar disk, and central protostar. Due to intense scattering and extinction in the dense envelope, direct imaging of the envelope is only possible in the millimeter regime. On the other hand, the incredible sensitivity of the Spitzer Space Telescope allows us to observe through the dense envelope enshrouding the young star to the scattered light within the outflow cavities. In the infrared, the outflow cavity is the most prominent feature \citep[e.g.][]{tob07}. Dust particles within the cavity scatter stellar and disk radiation into the line of sight, releasing light that would be completely obscured without the presence of the cavity. The shape and structure of the outflow can be determined by imaging this scattered light in the mid-IR. In this paper, we use archival \textit{Spitzer Space Telescope} data from 27 nearby YSOs to determine the opening angle of the outflow cavities. In order to apply the measured opening angles to an evolutionary scheme, we match observed photometry data to pre-computed model Spectral Energy Distributions (SEDs) of 20,000 YSOs at varying stages of prestellar evolution. We inserted mid-IR and far-IR data into a web-based\footnote{http://caravan.astro.wisc.edu/protostars/} SED fitting program \citep{rob07} to find the best fit models for the 27 nearby YSOs. Using the best-fit models, we identified evolutionary trends in the physical structure of young stellar objects and their surrounding environments.

\section{Observations and Photometry}

We used archived observations of nearby star forming regions in both the mid- (IRAC) and far-infrared (MIPS \& IRAS) continuum bands. The \textit{Spitzer Space Telescope} \citep{wer04} possesses two infrared cameras, the \textit{Infrared Array Camera}, IRAC \citep{faz04}, and the \textit{Multiband Imaging Photometer}, MIPS \citep{rie04}. IRAC, operating in the mid-IR, has four channels at 3.6, 4.5, 5.8, and 8.0 $\mu$m. MIPS, with three far-IR passbands, operates at 24, 70, and 160 $\mu$m. Also once operating in the far-IR is the \textit{Infrared Astronomical Satellite}, IRAS \citep{neu84}, sensitive to bands centered at 12, 25, 60, and 100 $\mu$m.

Archival data from these three cameras of nearby molecular clouds are used in our study. Molecular clouds are a breeding ground for stars, with the necessary raw materials and cold, dense conditions necessary for star formation. Although isolated YSOs within small dense clouds exist, a majority of forming stars are located within large molecular clouds. An inspection of these large molecular clouds is best suited for a manual search for a large sample of YSO candidates. The proximity of the observed YSOs is crucial for our study, as the outflow cavities must be resolved. Our search for YSOs consists of examining nearby (within 500 parsecs) molecular clouds. Many of the nearby molecular clouds of interest were observed by the \textit{Spitzer Space Telescope} (IRAC \& MIPS cameras) in the \textit{From Molecular Cores to Planet-Forming Disks} (c2d) \textit{Legacy Program} \citep{eva05}. Among the regions observed in this program were Chamaeleon II, Lupus, Ophiuchus, Perseus, and Serpens molecular clouds at distances of 200, 125, 125, 320, and 310 parsecs, respectively. Images of other nearby molecular clouds not covered by the \textit{Legacy} program (Orion, 450 pc; Cepheus, 180-450 pc; Taurus, 140 pc; and R coronae Australis, 170 pc) were obtained from the Spitzer archive\footnote{Data obtained via Spitzer's data delivery system \textit{Leopard}.}.

We performed aperture photometry on the IRAC and MIPS images for the 27 sources listed in Table 1 using the \textit{Image Reduction and Analysis Facility} (IRAF) program. Using the \textsf{phot} task within the \textsf{apphot} package, for each source, we summed the IRAC flux within an aperture physically corresponding to 500, 1000, 2000, and 4000 AUs. All four apertures were not possible for every source due to nearby stellar or ISM confusion. MIPS photometry was performed at a single aperture size for each source. Background subtraction is non-trivial since YSO outflow cavities are extended sources, eliminating the possibility for a background annulus. Instead, we used the \textsf{phot} task over a large non-stellar and non-ISM contaminated region to obtain an average background flux per pixel measurement. IRAC, MIPS 24 $\mu$m, and MIPS 70 $\mu$m photometry measurements are assumed to have a combined absolute calibration and measurement uncertainty of 10\%, 10\%, and 20\%, respectively mostly due to background subtraction difficulties \citep[e.g.][]{har05}. IRAS photometry was extracted from the \textit{IRAS Point Source Catalogue, 2nd Edition} available through the online SIMBAD database\footnote{http://simbad.u-strasbg.fr/simbad/}. When the flux measurement provided is not an upperlimit, IRAS measurements have an estimated uncertainty of 20\%. A full list of photometry measurements and apertures is given in Tables 3 and 4.

\section{Finding YSO Candidates}

We conducted an exhaustive visual search through some of the most well-known nearby star forming regions for YSO candidates. We used a visual search through the archive rather than focusing on existing catalogs of known YSOs for the following two reasons: (1) to allow for unknown sources and (2) to select those sources with the most obvious outflows. A visual search for both known and unknown YSOs is ideal for our purposes which required a large sample of YSOs with a resolved outflow cavity. Outflow cavities, an indicator of Class 0 and Class I protostars, are best observed in the mid-IR. Light scattered off dust particles within the cavity are best observed within our data at Channel 1, 3.6 $\mu$m, of the IRAC camera. We searched through a total of about 40 square degrees of archived IRAC data for outflows resulting from young stars. All of the cavities are identified by the common bow tie shape (if both outflow lobes are visible) or by the half bow tie shape (see Figure 1) when only one cavity is visible. The search resulted in the visual identification of 38 nearby YSOs, four of which were previously unknown sources. 

Our strategy for identifying sources resulted in a biased catalog of YSOs. The identified YSOs all must possess an outflow cavity, and that cavity must be fairly luminous in order to be visually recognized. Additionally, all of the YSOs in the sample must have fairly high line-of-sight inclination angles (nearly disk edge-on). Low inclination, near pole-on, inclination angle YSOs cannot be identified visually since the classic bow tie shape will not be present when looking down the ``throat'' of the outflow. All YSOs in our catalog must be inclined by less than half the full outflow opening angle. This inclination bias may have the effect of excluding YSOs with large opening angled outflows, since they have a greater chance of being viewed down the cavity. Not all of these 38 sources are ideal for our study. Some outflow cavities lack a measurable structure due to low luminosity, indefinite cavity walls, or small angular size. Additionally, some sources are not optimal for photometry. Molecular clouds can contain complex, bright ISM structures like filaments that can cause confusion when doing photometry. Ultimately, there are 27 YSOs with clear, quantifiable outflows that provide reliable photometry. Two of these YSOs in the final catalog of 27, SL 05346$-$0528 and SL 18299+0117, were previously uncataloged sources according to the online astronomical database SIMBAD. A full list of the 27 YSOs used in the study is provided in Table 1 along with positions and estimated distances. For the two newly discovered objects, the distances are assumed to be the average distance to the molecular cloud in which it is located. For use in the SED model fitting, all distance uncertainties are assumed to be 20 pc.

\section{Determining Cavity Opening Angles}

As can be seen in the IRAC channel 1 images of the YSOs (Figure 1), the walls of the cavity show prominent limb brightening. These walls define an edge and therefore a measurable size to the cavity. In order to quantify the opening angles of the outflow cavities, we made annular flux cuts around the YSOs (Figure 2). For consistency, we made our circular cuts on IRAC channel 1, 3.6 $\mu$m, data only. The cavity walls are often less defined in the longer IRAC wavelengths, preventing a reliable edge from being determined. As shown in Figure 2, the infrared flux from the cavity peaks at the cavity walls. 

We define the opening angle of each YSO cavity to be the angle between the high-points in the annular cuts. The radius of each annular cut was varied to the largest possible value which would still show two prominent intensity peaks at the cavity edges. The measurement of the opening angle is based upon this single annular cut at the measureable extent of the outflow cavity. We find that so long as the radius of the cut is suffiently large, the measurement of the opening angle is independent, within the uncertainty of the measurement, of the annular radius. While at the base of the cavities the walls may be curved, at large distances from the central source, the cavity walls become largely straight and radial, removing the dependence of opening angle on radius. Each cut has an annular thickness of one pixel, and the center of each flux annulus is the brightest point at the base of the cavity. It is important to note that this central and brightest point is most likely not the protostar itself, but rather the brighter, denser ``throat'' of the outflow. Because of the selection bias towards high-inclination YSOs, for all of our sources, the protostar itself is obscured by the envelope.

Superimposed on Figure 1 are dotted lines indicating the measured locations of the YSOs cavity walls as determined by the annular cuts. Each pair of walls defines a cavity opening angle, and in Table 2, we list the measured opening angles for each of our YSO candidate sources. Our catalog of sources contains YSOs with only one cavity lobe and both cavities being visible. When both cavities are visible, it is common for the opening angle of one cavity to be slightly inconsistent with the other. We have measured and recorded the opening angles for all visible cavities, and have listed the maximum and minimum values for each source in Table 2 along with the average of the two. When only one cavity is visible, the average value is simply the measured opening angle of the one visible side.

When only one outflow cavity is visible, the ``backside'' cavity has been obscured beyond detection by high extinction from the circumstellar envelope. Although extinction can cause one cavity to be unobservable, it does not seem to have an effect on the measured opening angle of the remaining visible cavity. Since the cavity walls are highlighted by limb brightening, the presence of peaks in the circular flux cuts signals the presence of a cavity wall, a quality unchanged by excess extinction. Model YSO images created using the radiation transfer code of \citet{whi03a,whi03b} confirm that increasing and decreasing the envelope density, and therefore mass, will only affect the measured cavity opening angles within the uncertainty of the angle measurement.

The opening angle measured directly from the IRAC images is dependent on both the true opening angle and the inclination at which the object is being viewed. Only when the objects are being viewed edge-on (defined as a 90$\arcdeg$ viewing angle) are the true and measured opening angles the same. Due to this effect from inclination viewing angle, the value of the opening angle measured directly form the IRAC image is truly only an upper limit. In an attempt to correct for the effects of inclination, we determine a best-fit viewing angle through 2-D radiation transfer computer modeling (see section 5) and work backwards to determine the true, edge-on, cavity opening angle. A simple geometric relation $\phi=2\arcsin(\sin(\frac{\phi'}{2})\sin(\theta))$ where $\phi$ is the true cavity opening angle, $\phi'$ is the measured opening angle, and $\theta$ is the angle of inclination, is used to determine the true opening angle. Here, pole-on is defined as 0$\arcdeg$ of inclination and 90$\arcdeg$ is edge-on. Note that any inclination angle less than 90$\arcdeg$ results in a true opening angle smaller than the measured one. Because our catalog is biased towards high-inclination YSOs, the difference between measured and actual cavity opening angles is quite small for most sources.

\section{Modeling YSOs}

\subsection{Method}

An important yet difficult part of observational studies of YSOs is the identification of the evolutionary states of young sources \citep[e.g.][]{ada87}. Since it is not possible to observe a single YSO as it ages over a significant evolutionary time scale, all observational studies must be done by observing many YSOs, and from them, inferring an evolutionary sequence. Here, we use multi-wavelength photometry to construct SEDs of the YSOs, and infer an evolutionary state directly from the SED. To analyze the SEDs, we compare them to pre-computed model SEDs constructed from 2-D radiation transfer models in a web-based fitter \citep{rob07}. Ultimately, we wish to find a combination of physical parameters in the models that produces an SED that matches the SEDs of the observed YSOs. While other techniques (color-color diagrams, spectral indices) can be useful in identifying evolutionary stages, this technique has the unique ability to infer physical information about the young star such as stellar mass, accretion rate, and cavity density. Although SED fits have a notorious heritage, it is important to note that in this case we are evaluating evolutionary trends, so we are interested in the relative YSO parameters.

The model SEDs used are a result of a 2-D Monte Carlo radiation transfer code developed by \citet{whi03a,whi03b,whi04}; we use the web-based SED fitting interface\footnote{http://caravan.astro.wisc.edu/protostars/index.php} developed by \citet{rob07}. The model YSOs consist of a central protostar, rotationally-flattened infalling envelope, bipolar cavities, and flared accretion disk. There are a total of 14 model parameters that are sampled within a range to create a total of 200,000 model SEDs. The full list of parameters is given in Table 1 of \citet{rob06}. These 14 parameters are arranged into 20,000 combinations with ten possible inclinations (from pole-on to edge-on in equal intervals of cosine of the inclination) for each set of parameters, giving a total of 200,000 model SEDs. Not all 14 model parameters have a large effect on the resulting SED at a given evolutionary stage. In the earliest prestellar stages of which we are interested, the most important parameters are the envelope accretion rate, opening angle of the bipolar cavities, line-of-sight inclination angle, disk and envelope inner radius, stellar temperature, and disk mass \citep{rob07}. It is important to note which parameters have the strongest effect on the SED in order to avoid any over analysis of the matching parameters. 

The grid of 20,000 model YSOs covers a large range of parameter space, allowing for as few assumptions about evolutionary changes as possible while still keeping computational time to a minimum. The parameter ranges span those predicted by both theoretical and observational work. The parameters of the protostar (stellar mass, radius, and temperature), the infalling envelope (envelope accretion rate, outer and inner radii, cavity opening angle, and cavity density), and the accretion disk (disk mass, accretion rate, outer and inner radii, flaring power, and scale height) can all be varied. The stellar masses are sampled between 0.1 and 50 solar masses, and prestellar ages can range from $10^{3}$ to $10^{7}$ years. Using evolutionary tracks \citep{ber96, sie00}, each pair of stellar mass and age can be used to determine the stellar radius and temperature. The remaining disk and envelope parameters are randomly sampled from ranges that depend on the age of the source and are supported by observational and theoretical work. For details about the evolutionary ranges of the parameters, refer to \citet{rob06}.

We used data from IRAC, MIPS, and IRAS to construct SEDs for the 27 nearby YSOs in our catalog. In an attempt to better constrain the output parameters for each YSO, we used several aperture sizes for each waveband in the IRAC data as a poor-man's image fitting technique. This allows the fitting tool to obtain some elementary geometrical data about each YSO. This technique is only possible for wavebands in which the YSO is resolved, and is therefore only possible for IRAC images. Though it not always possible to use all four apertures (due to nearby stellar or interstellar medium confusion), it is important to use as many apertures as possible to better constrain the models. Since there are four IRAC bands, there a total of 16 possible data points from IRAC data alone. The inclusion of MIPS and IRAS data is vital to the determination of the evolutionary state of the YSOs; IRAC data alone may not provide ample information to construct an accurate SED. Though commonly used, IRAC colors alone are not a good indicator of evolutionary stage. For the youngest sources, radiation transfer models of still-forming stars find that the mid-IR SED shape, and therefore color, has in important dependence on the orientation of the disk and envelope relative to the line-of-sight \citep{whi03a}. Low inclinations (pole-on viewings) are generally bluer than higher inclination (edge-on viewing) sources. In order to separate evolutionary changes from inclination effects, the inclusion of longer-wavelength data is necessary to more accurately match SEDs and infer an evolutionary stage. Tables 3 and 4 show the flux density, when operative, from each of the 22 possible points (4 IRAC bands at 4 aperture sizes, 2 available MIPS bands, and 4 IRAS bands). 

\subsection{Modeling Results}

Shown below in Figure 3 are the SEDs for all 27 of our sources along with the best fit model for each source. A measure of the quality of the fit, reduced chi-squared per data point, was determined for each source-model pair. Typically, when comparing a grid of models to observations, it is useful to define a chi-squared cut-off level that corresponds to a particular confidence level (80 or 90$\%$, for example). In this case, the grid does not allow an exploration of parameter space, so one doesn't necessarily expect a good fit. SED fitting of this nature is not optimized for chi-squared minimization. Since it is difficult, if not impossible, to determine if the physical components of the models are consistent with the properties of the true YSOs, we argue that a chi-squared cut-off is statistically impossible to define \citep[also see][]{rob07}. Other statistical complications include intrinsic stellar variability and asymmetry in the protostar-disk-envelope system that is unaccounted-for in the computer models. In order to evaluate evolutionary trends in our 27 YSOs, we will accept the best fit model as an evolutionary match; we argue that our large sample of sources will compensate for misidentifications. There is a large range of chi-squared values in our sample: the minimum reduced chi-squared is 28.31, and the maximum reduced chi-squared is 8616.4. With an average of 761.47, and a median of 210.01, we argue that the models, in general, do an adequate job of matching the data. Only 3 sources (11$\%$) have a chi-squared above 777. Almost half the sources have a chi-squared below 200, and 70$\%$ of the sources have a chi-squared of less than 400.

Along with providing estimated ages, cavity densities, etc., each model fit provides an estimate for the line-of-sight inclination angle. Since viewing angle has an effect on the projected, measured cavity opening angle, it is necessary to decouple inclination effects from the true opening angles. Using the model-estimated inclinations, we corrected each average measured opening angle as descbed in Section 4 to determine what the true, edge-on cavity opening angle is for each source. The inclination-corrected average opening angles for each source are provided in Table 2. Since the fitter is only provided with circularly averaged photometry, we do not expect the SED fitter to accurately predict the true opening angle of the cavity. Because of the strong effect inclination has on the shape of mid-IR portion of the SED, we believe the inclination to be a more robust estimation, so we argue that the cavity angles measured directly from the images and corrected by the model inclinations are more accurate representations of the true opening angles than the modeled cavity opening angles. It is possible the fitter is not accuratley determining the inclination angles, in which case the corrections we are applying to the opening angles adds random noise to the measured values. As noted in Section 4, because the inclination angles are determined to nearly all be about 90\arcdeg, any noise would be small. We will accept the corrected opening angle values since at best the correction is accurate, and at worst, the correction adds only a small amount of noise.

\section{Discussion}

\subsection{Variation of Cavity Opening Angle With Age}

The outflows in Figure 1 and their circular cuts in Figure 2 show that there are a variety of morphologies for YSOs. We argue that there are trends within the array of morphologies due in part, if not entirely, to the evolution of the protostar. In order to assess evolutionary changes in the cavity morphologies, we explore the image-measured, model-inclination corrected cavity opening angles. In Figure 4 we plot the average inclination-corrected opening angle of the outflow cavity as a function of each source's estimated age as fit by the SED. It is clear that there exists a correlation between evolutionary stage and outflow opening angle $-$ with the angle tending to increase with time. The correlation is positive, with a Pearson product-moment correlation coefficient of 0.34. Since the correlation between age and angle may not be linear, Kendall's Tau rank correlation coefficient may be a more robust measurement of the correlation strength. Using Kendall's Tau, we determine a probability of no correlation of 19$\%$. Pearson product-moment correlation coefficients with associated probabilities, Kendall's Tau rank correlation coefficients, and the probabilities of no correlation for Figures 4$-$11 are provided in Table 5. A linear fit to our data yields $\log(\theta/$deg$)=(1.3\pm0.1)+(0.11\pm0.01)\log(t/$yr$)$. This is consistent with the linear fit $\log(\theta/$deg$)=(1.1\pm0.2)+(0.16\pm0.4)\log(t/$yr$)$ determined by \citet{arc06} from CO observations of a different sample of YSOs. An additional analysis by \citet{arc06} combined their YSO sample with sources from the literature to find a combined data fit of $\log(\theta/$deg$)=(0.7\pm0.2)+(0.26\pm0.4)\log(t/$yr$)$, which is consistent with our findings. The true uncertainty of our fit parameters may in actuality be larger; the relatively small quoted uncertainty is attributed to an assumption of no age estimation uncertainty. 

The classification system described in Section 1 (Class 0, I, II, III) is a widely used scheme, and historically, YSOs are grouped into these classes based on color, spectral index, or luminosity ratio. But because of the profound effect viewing inclination can have on these parameters, this classification scheme may not be the best method to describe prestellar evolution. Some protostars can be classified in two, or possibly three, classes depending on the classification method used \citep{ken93a,ken93b,yor93,son95,whi03a,whi04}. For example, \citet{whi03a} showed that moderately-inclined (40\arcdeg) Class I protostars have optical, near- and far-infrared fluxes similar to more edge-on (75\arcdeg) Class II sources. The color effects due to inclination are the most profound in the near- and mid-infrared regimes \citep{whi07}. 

To evidence the pitfalls of the class scheme, we have determined the classes of the 27 YSOs by two methods. Both originally developed by \citet{and93}, they use specific millimeter and submillimeter spectral information to estimate class. A Class 0 YSO was defined as a source yet to accrete much material onto the central source, so was conceptually defined as having $M_{star}/M_{env}$ $<$ 1. Given some assumptions about mass-infall rate, dust opacity, and dust temperature, this roughly corresponds to $L_{bol}/L_{1.3}$ $\leq$ $2 \times 10^{4}$ where $L_{1.3}$ is the luminosity at 1.3 mm and $L_{bol}$ is the bolometric luminosity. This definition roughly corresponds to $L_{submm}/L_{bol}$ $\geq$ $5 \times 10^{-3}$ where $L_{submm}$ is the luminosity radiated longward of 350 $\mu$m. Given that these class divisions are based on assumptions and simplified circumstellar models, the cut-offs between Class 0 and Class 1 are vague, and somewhat arbitrary, with many sources lying very close to the class boundaries. This should be expected since these is a continuity between Class 0 and Class 1. We therefore define a Class 0/I, which lie close to the class boundaries. For the 1.3 mm test, we define Class 0 as objects that have $L_{bol}/L_{1.3}$ $\leq$ $2 \times 10^{4}$, Class 0/I objects as those with $L_{bol}/L_{1.3}$ $\leq$ $8 \times 10^{4}$, and Class I Objects as those with $L_{bol}/L_{1.3}$ $>$ $8 \times 10^{4}$.  For the submillimeter test, we define Class 0 as objects that have $L_{submm}/L_{bol}$ $\geq$ $10^{-2}$, Class 0/I objects as those with $L_{submm}/L_{bol}$ $\geq$ $5 \times 10^{-3}$, and Class I objects as those with $L_{submm}/L_{bol}$ $<$ $5 \times 10^{-3}$. Again, the deliniations are somewhat arbitrary, but provide a general evolutionary scheme. For each of the catalog YSOs, we determined the class from both methods by using the SEDs produced by the fitting tool. The results of the classification are in Table 1. The two categorization schemes show general agreement, but also highlight the problem with classifications based solely on spectral measurements; 4 of the 27 YSOs were determined by the 1.3 mm test to be Class I but were categorized as Class 0 by the submillimeter test.

We also adopt a taxonomy scheme developed by \citet{rob06} that classifies stars in an evolutionary fashion as opposed to using SED qualities. YSOs age from Stages 0 to III where Stage 0 and I objects (here combined into a single stage, Stage I) are characterized by dense infalling envelopes and possibly disks. Stage II objects have optically thick disks and a tenuous envelope, and Stage III YSOs have optically thin disks. We use the output parameters of the SED fitting tool in order to classify the 27 sources into each of these categories. Stage I objects have $\dot{M}_{env}/M_{star}$ $>$ $10^{-6}$ year$^{-1}$. Stage II objects are those with $\dot{M}_{env}/M_{star} $ $<$ $ 10^{-6}$ year$^{-1}$ and $M_{disk}/M_{star}$ $>$ $10^{-6}$, while Stage III objects are those with $\dot{M}_{env}/M_{star}$ $<$ $10^{-6}$ year$^{-1}$ and $M_{disk}/M_{star}$ $<$ $10^{-6}$. The exact boundaries between stages are, as with class, arbitrary, but should aide in giving a ``feel'' for the general evolutionary trends. Since all of our YSOs have cavities and envelopes, only Stage I and II objects should be expected to be present in out sample. This is indeed the case, and these two general stages are differentiated in Figures 4$-$11 as follows: YSOs classified as Stage I are marked with filled squares and Stage II with open ones. Table 1 provides the determined stage for each YSO. Note that there is generally a correlation betwen the stage and class systems. All Stage I sources were identified as Class 0 YSOs by at least one of the class tests. Additionally, any YSO determined by one test to be a Class I and by the other to be a Class I or Class 0/I was identified as a Stage II source. Note from Figure 4 that the 16 oldest-esimated YSOs are all Stage II objects, with a Stage I object never exceeding an age of $10^{4.3}$ years. In our sample, Stage I and Stage II objects have similar average and median opening angles (identical within uncertainty), but as can be seen in the following figures, the outflow cavity angles have the general trend of increasing as the YSOs age from Stage I to Stage II. The six largest opening angles belong to cavities resulting from Stage II objects.

Since both age and cavity opening angle are output parameters for the model SED fitter, one possible concern with the validity of the age-opening angle relation is that there is an innate correlation built into the SED fitting code to cause this relationship. The array of 20,000 model YSOs range in age from $10^{3}$ to $10^{7}$ years. An individual YSO at a given age is randomly assigned a cavity opening angle within a particular angle range, and the average value of this range increases with source age. Therefore it is inherent within the code that YSOs with the smallest opening angles are also the youngest. For this to have an effect on the age-measured opening angle relation, the fitter would have to accurately determine the opening angle of the cavity outflow. Figure 5 is a plot of the measured, inclination-corrected outflow angle as a function of the opening angle determined by the fitter program. Although there may be a small correlation (the three sources with the smallest measured opening angles also have the smallest fit opening angles), there exists no strong correlation that would cause the age-angle relation in Figure 4 to be caused by the SED fitter. As an example, IRAS 03292+3039 has the largest opening angle measured directly from the IRAC images in our sample (82$\arcdeg$), but is assigned a relatively small opening angle of under 30$\arcdeg$. With a Pearson correlation coefficient of 0.19 and a probability of no correaltion of 63\% clearly, the SED fitter does a poor job of determining the cavities' opening angles. As concluded by \citet{rob07}, the fitter tool's inability to accurately determine the opening angle should be expected $-$ the fitting tool is only provided with circularly averaged flux measurements that lack any geometrical information.

\subsection{Variation of Cavity Opening Angle With YSO Physical Parameters}

Since the age estimates determined by the fitter may not be accurate, it may be useful to identify evolutionary effects by using physical characteristics of the YSOs as evolutionary indicators. A common indicator of YSO evolutionary stage is the density within the outflow cavity. As the protostar ages, and the outflow clears the surrounding envelope, the density within the cavity decreases; note in Figures 6 and 7 that the 10 objects with the lowest estimated cavity densities are all Stage II YSOs. Figure 6 plots the fit cavity density as a function of fit YSO age. Note that the strong correlation is in part, if not fully, a product of the bias within the grid of 200,000 model YSOs. The range of values from which the density of gas and dust within the cavity was sampled followed a deceasing function of time. Therefore, as a YSO ages, the density of material within the cavity necessarily decreases. Should this be an accurate picture of star formation, it should then follow that the outflow cavities with the highest densities should also have the smallest opening angles. Figure 7 plots the measured opening angle corrected for inclination as a function of the cavity density as determined by the SED fitting tool. Although the correlation with cavity density is not as tight as with age, there does appear to generally be an inverse relation. As a general trend, the lowest density cavities, and therefore the most evolved, have the largest measured cavity opening angles. The weakness of the correlation may be due to the following fact: although the cavity density of a single YSO will decrease in time, when comparing cavities, a lower density cavity may not necessarily belong to a more evolved YSO. Cavity densities are not purely evolutionary and result from initial envelope densities and complex outflow dynamics. 

Another potential evolutionary indicator for young stars is the accretion rate. Figure 8 plots the fitted accretion rate as a function of fitted YSO age, and demonstrates that as a protostar ages and loses envelope material to infall or outflow, the accretion rate should decrease. Figures 8 and 9 show that only one of the 22 Stage II YSOs has a modeled accretion rate higher than a Stage I YSO. Again, the correlation within the figure is in part a product of model bias. The accretion rate was sampled from a range of values that was constant for ages $<10^{4}$ yr, and decreases with times thereafter, going to zero at about $10^{6}$ yr. Figure 9 plots the inclination-corrected measured cavity opening angle against accretion rate. Like the cavity density, the figure shows a general correlation - the YSOs with the lowest accretion rate, and therefore the most evolved, tend to have the largest cavity opening angles. In fact, according to the models, with the largest cavity opening angle in our catalog, IRAS 03292+3039 has almost completely ceased accreting. This correlation is the weakest of the angle-parameter correlations (Figures 4, 7, 9) $-$ a selection of 27 points from a random distribution of angles and accretion rates would have an equal stronger Pearson product-moment correlation coefficient 18\% of the time. The weak correlation is also evidence by the very high probability (49\%) of no correlation determined by Kendall's Tau

\subsection{Color as an Indicator of Age}

A common practice in the study of YSOs is to use various photometric colors in order to identify evolutionary changes. With the launch of the \textit{Spitzer Space Telescope} and its IRAC camera, the use of mid-IR color-color and color-magnitude plots has become widely used to classify YSOs. \citet{rob06} showed that these mid-IR  (such as IRAC) data can be an effective way of separating YSOs from main sequence stars. Pre-main sequence stars tend to lie redward in IRAC's [3.6]-[4.5] and [5.8]-[8.0] colors of stellar photospheres, which tend to cluster around (0,0). Additionally, they conclude that some of the youngest YSOs may be identifiable using IRAC data only. Stage I YSOs, the least evolved sources, occupy a large region of color-color space that is unoccupied by Stage II and Stage III YSOs. But there are also areas in color-color space occupied by all three stages. Given this, it is difficult to use mid-IR photometry alone to identify evolved and highly-evolved sources as such. As a test of the efficiency of color-identified evolution, we did photometry on our 27 nearby resolved YSOs. We used the fluxes integrated in a circular aperture corresponding to a radius of 2000 AU around the bright base of the outflow cavity. Figure 10 shows the evolutionary changes of various IRAC colors as a function of age as determined by the fitter tool for our 27 sources. The youngest estimated sources do tend to lie redward of the older sources in the [3.6]-[4.5], [3.6]-[5.8], [3.6]-[8.0] colors. A best fit line has been plotted for these three colors. Although a slight negative correlation does exist between the [4.5]-[5.8], [4.5]-[8.0], and [5.8]-[8.0] colors, the correlation is so weak, a fit line is statistically meaningless. A study by \citet{ale04} shows that the [3.6]-[4.5] color is a good indicator of envelope density, with the highest densities corresponding to the youngest YSOs. This is generally consistent with our color-age plots. The estimated youngest YSOs, which have newly-formed outflows that have yet to clear much envelope away, are generally the reddest of our sources. The weakest age-color correlation involves the [5.8]-[8.0] color. The [5.8]-[8.0] color is suspected to be an indicator of stellar luminosity, and therefore mass \citep{ale04}. Since stellar mass is evolutionarily uncorrelated with YSO age, it should not to surprising that there is no clear correlation between [5.8]-[8.0] color and age.

Figure 11 plots the measured, inclination corrected cavity opening angles as a function of IRAC color. If the reddest sources are the youngest, then we should expect the cavity opening angles to be the smallest for the reddest YSOs. Indeed, all the color vs. opening angle plots show a negative correlation. Since age does not appear to be correlated with [4.5]-[5.8], [4.5]-[8.0], or [5.8]-[8.0], as indicated by the small correlation coefficients in Figure 10, any correlation between color and opening angle in these colors may not be due to evolutionary effects. Note that even in the colors associated with evolutionary stage ([3.6]-[4.5], [3.6]-[5.8], and [3.6]-[8.0]), the bluest sources do not necessarily have the largest cavity opening angles. This is not unexpected since IRAC colors do a poor job of identifying more evolved (Stage II) YSOs. These plots indicate there may be a correlation between color and cavity opening angle, but examination of Figures 8$-$9 shows that using only mid-IR photometry is a poor method of identifying evolutionary stages. Although using IRAC data can be effective for some YSOs, the addition of longer wavelength data is necessary to accurately identify the evolutionary stage of the young stars. \citet{rob06} suggests that the addition of photometry for wavelengths beyond about 20 $\mu$m (such as that provided by MIPS and IRAS) substantially enhances the ability to determine the evolutionary stage of YSOs. SED fitting, as used in this paper, is arguably the best and most effective use of available photometric data since it allows for the inclusion of a wide range of wavelength data.

\subsection{Comparisons with Molecular Outflows}

Although the age-cavity opening angle correlation identified in this paper has been implied by other authors, and is expected to exist under current theoretical models, it has never before been revealed through the use of SED-model fitting. High-resolution images by \citet{vel98} implied that the cavity opening angle of B5-IRS1 is widening at a rate of 0$\arcdeg$.006 year$^{-1}$. Using CO data to map molecular outflows, \citet{arc06} showed that cavity opening angle widening is a more general trend of YSOs. In that study, source age was determined using the $T_{bol}-$age relation from \citet{lad98}. Combined with prior studies such as these, our work clearly demonstrates that outflow opening angles increase as sources age.

Several of the YSO outflows in our sample have been studied with molecular line observations. Measurements of the outflow opening angle determined by CO measurements are also provided in Table 2. Of the three sources with measured opening angles from both scattered mid-IR and CO observations, two of them are substantially wider when measured in CO. The third outflow is, within uncertainty, identical in shape. We argue that the wider outflow from CO observations is due to the entrainment of material from the envelope into the cavity. Mid-IR observations highlight the walls of the carved-out cavity, but material just beyond the wall is being swept into the cavity by the jet/prestellar wind. We propose that the observed wider CO component belongs to the material being entrained into the cavity just beyond the cavity wall.

\section{Conclusion}

Our thorough search through nearby star forming regions has resulted in a sample of resolved YSOs and their companion outflow cavities that enables a detailed study of the morphological evolutionary effects to outflows. Estimated ages of the catalog sources have been explored through the use of IR-color and SED fitting to pre-computed models. The latter allows an extrapolation of relative physical parameters such as cavity density and accretion rate.

The mid-IR images provided by the IRAC camera give evidence to the presence of a cavity that has been cleared of its dense envelope material by an outflow. There appears to be an evolutionary trend in the morphology of the outflow cavities. The sources estimated to be the youngest by SED fitting and color possess the narrowest cavities. More evolved sources generally have larger, more open cavitites, indicating a widening of outflow cavities with time. Should our conclusions be correct that outlfow cavity angles widen over time, our work should provdie confindence that the 2-D radiation tranasfer models do, to first order, accuratley depict the structure and physics associated with star formation. Outflows are clearly affected by the aging of young pre-main sequence stars, and it is likely they play a major role in the star formation process. A detailed knowledge of the morphological changes to a cavity should produce insights into the outflow's creation and evolution, ultimately providing more accurate theories of star formation.

\acknowledgments
The authors wish to thank T. P. Robitaille, B. A. Whitney, and R. Indebetouw for the use of their models and SED fitter along with helpful discussions regarding its use, Robert Gruendl for valuable discussions and creation of the annular flux cuts tool, Woojin Kwon and John Tobin for early helpful discussions, and the anonymous referee for comments which greatly improved the quality and strength of this paper. This research has made use of the SIMBAD database, operated at CDS, Strasbourg, France. This work is based on archival data obtained with the \textit{Spitzer Space Telescope}, which is operated by the Jet Propulsion Laboratory, California Institute of Technology under a contract with NASA. This publication makes use of data from the \textit{Infrared Astronomical Satellite}. Support for this work was provided by NASA.

{\it Facilities:} \facility{Spitzer (IRAC, MIPS)}, \facility{IRAS}

\clearpage

\begin{figure}
%%1
\epsscale{.9}
\plotone{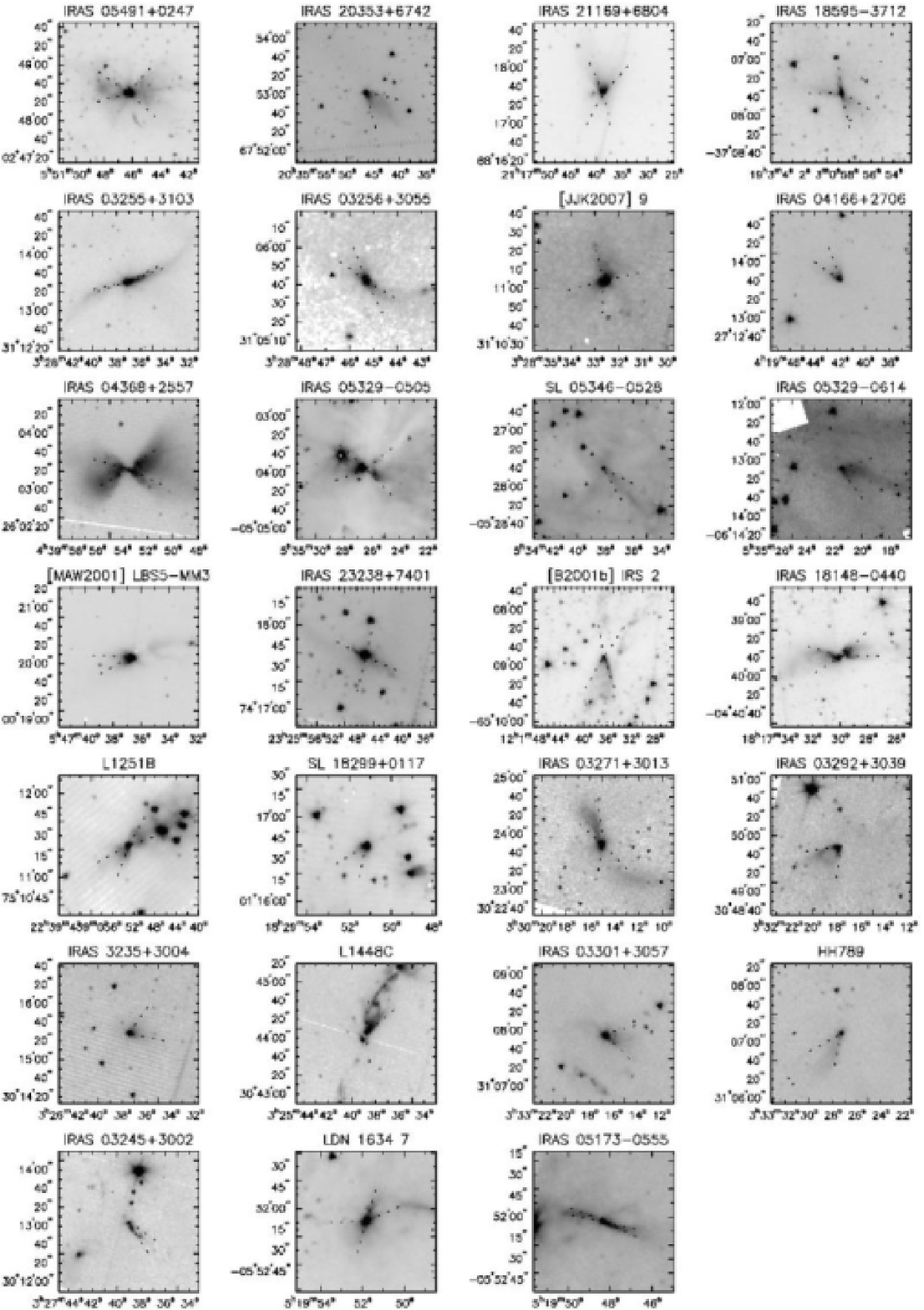}
\caption{Square root scaled IRAC channel 1 images of the 27 young stellar objects in the catalog. Dotted lines indicate the measured locations of the outflow cavity walls according to the angles determined by the annular cuts.}
\end{figure}

\clearpage

\begin{figure}
%%2
\epsscale{.7}
\plotone{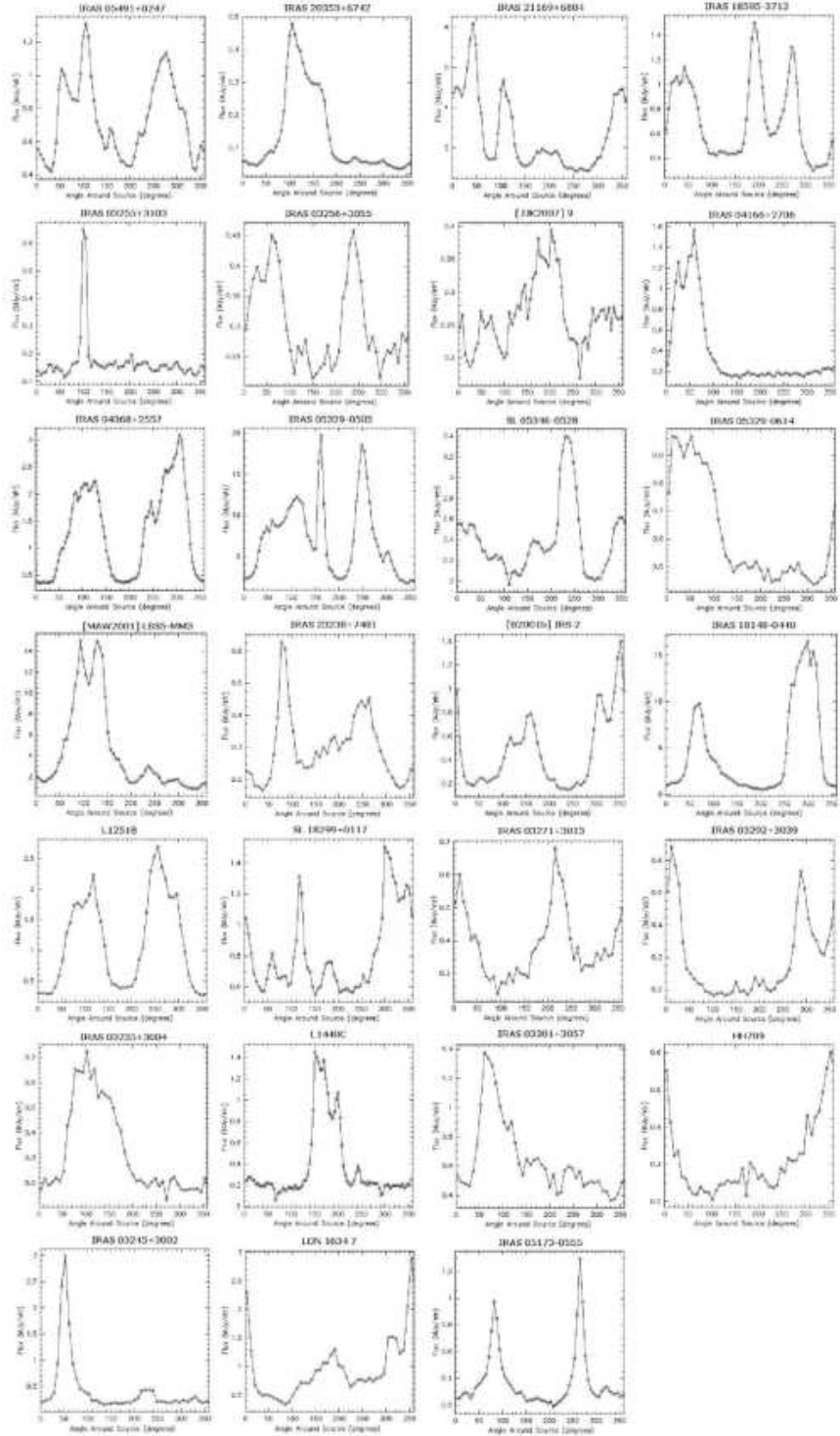}
\caption{Flux density vs. angle around source for the 27 young stellar objects in the catalog. The annular cuts are taken around the channel 1 IRAC images.}
\end{figure}

\clearpage

\begin{figure}
%%3
\epsscale{.8}
\plotone{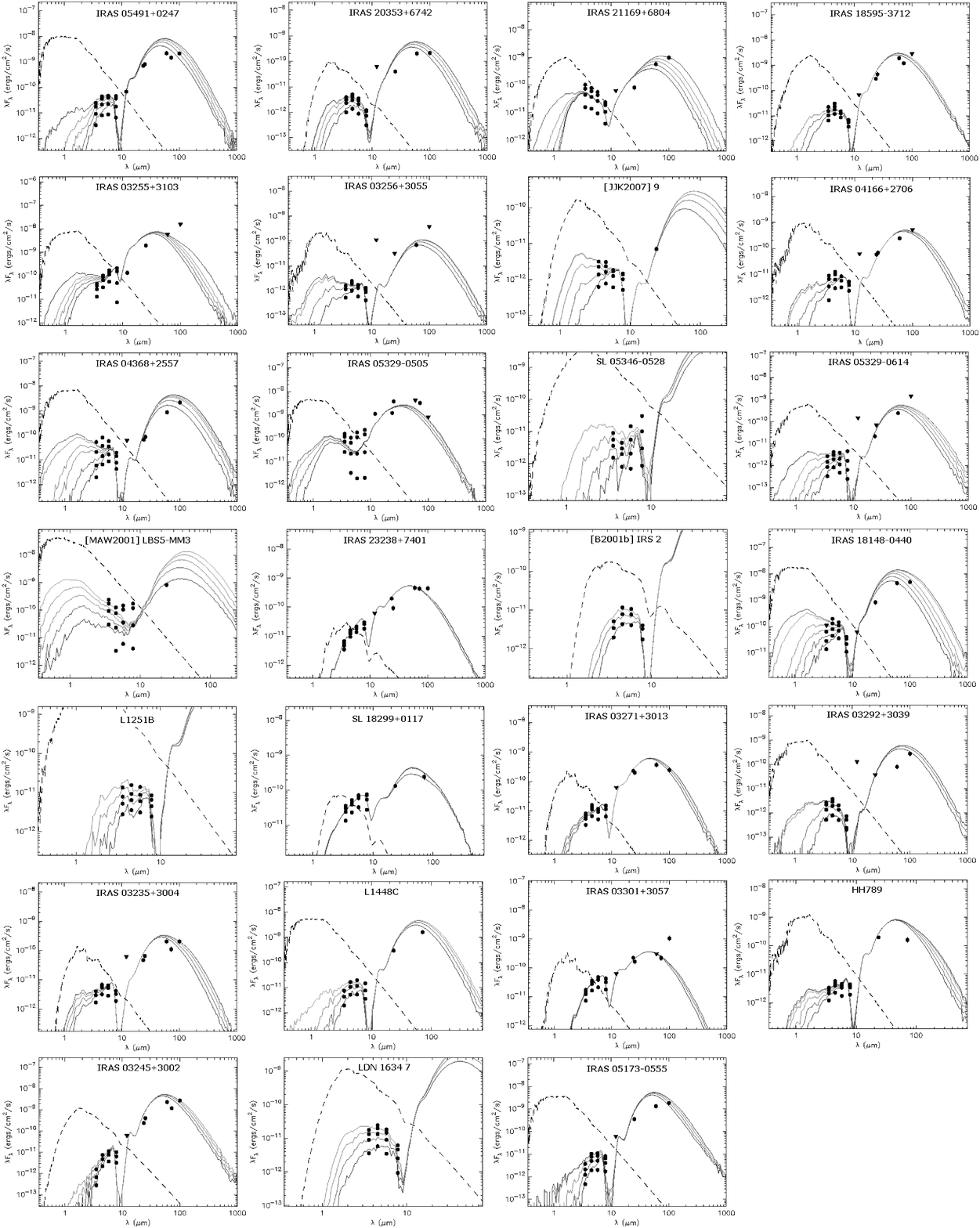}
\caption{Aperture-dependent model SEDs with measured photometry for each of the 27 YSOs in the catalog. Filled squares mark flux points, while triangles represent flux upperlimits. The solid curves are the best-fit model SEDs at all the apertures used in the data set. Apertures used for each YSO are given in Tables 3 and 4. The dashed line shows the stellar photosphere model that was used as input to the radiation transfer code for the best fit model.}
\end{figure}

\clearpage

\begin{figure}
%%4
\epsscale{.6}
\plotone{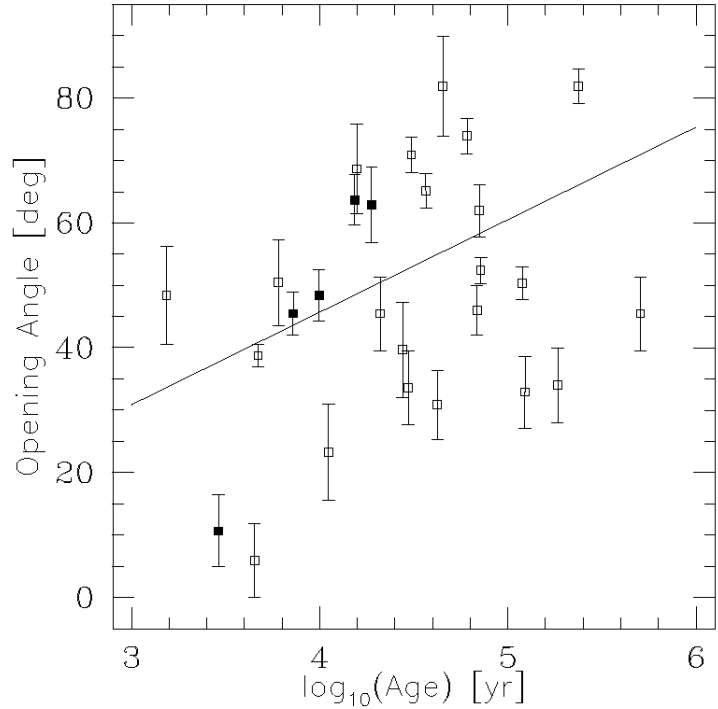}
\caption{Measured outflow opening angle as a function of modeled source age. The YSO ages are determined by the best-fit model SED. Opening angles are measured directly from the IRAC channel 1 images and are corrected for model inclination. The uncertainties in the opening angles are from image measurement. Filled in squares show sources in our sample that are catagorized as Stage I. Open squares represent objects that are Stage II YSOs. The line represents the linear fit to the data. }
\end{figure}

\clearpage

\begin{figure}
%%5
\epsscale{.6}
\plotone{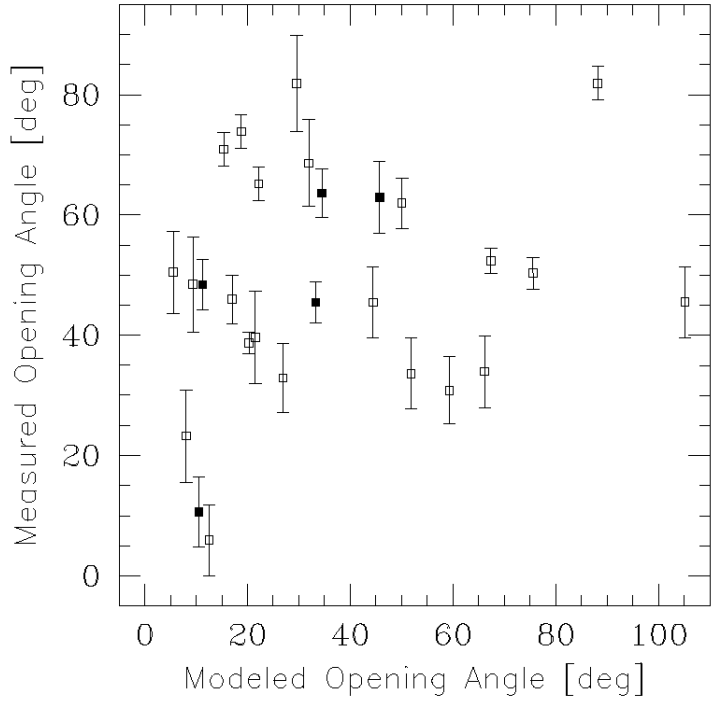}
\caption{Measured outflow opening angle as a function of modeled cavity opening angle. The estimated opening angles were determined by the best-fit model SED. Measured opening angles are taken directly from the IRAC channel 1 images and are corrected for inclination. The uncertainties in the opening angles are from image measurement. Filled in squares show sources in our sample that are catagorized as Stage I. Open squares represent objects that are Stage II YSOs. }
\end{figure}

\clearpage

\begin{figure}
%%6
\epsscale{.6}
\plotone{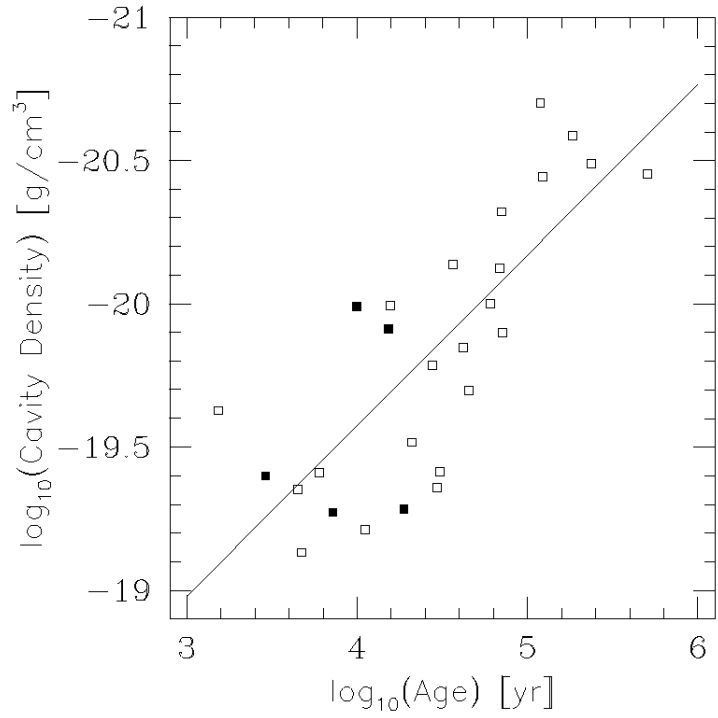}
\caption{Modeled cavity density as a function of modeled age. The YSO ages and cavity densities are determined by the best-fit model SED. Filled in squares show sources in our sample that are catagorized as Stage I. Open squares represent objects that are Stage II YSOs. The line represents the linear fit to the data. }
\end{figure}

\clearpage

\begin{figure}
%%7
\epsscale{.6}
\plotone{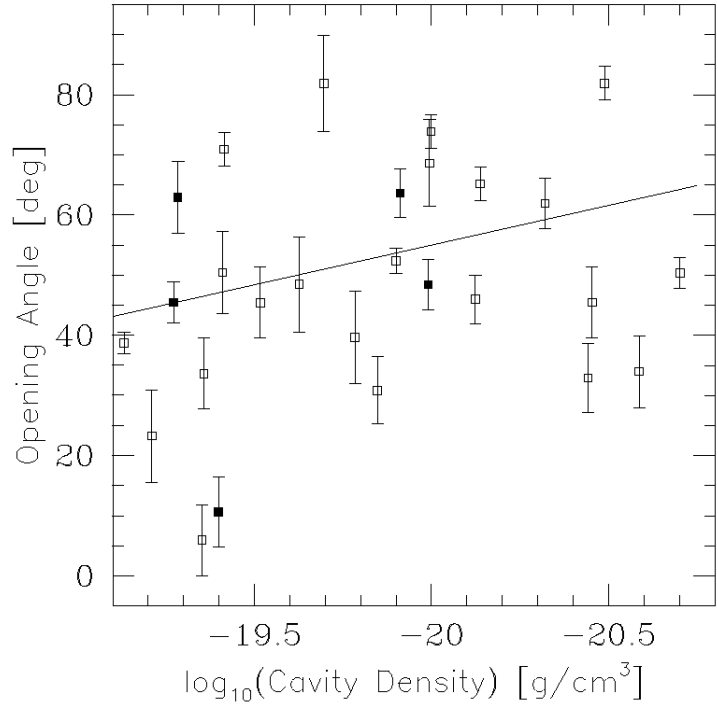}
\caption{Measured outflow opening angle as a function of modeled cavity density. The YSO cavity densities were determined by the best-fit model SED. Opening angles are measured directly from the IRAC channel 1 images and are corrected for inclination. The uncertainties in the opening angles are from image measurement. Filled in squares show sources in our sample that are catagorized as Stage I. Open squares represent objects that are Stage II YSOs. The line represents the linear fit to the data. }
\end{figure}

\clearpage

\begin{figure}
%%8
\epsscale{.6}
\plotone{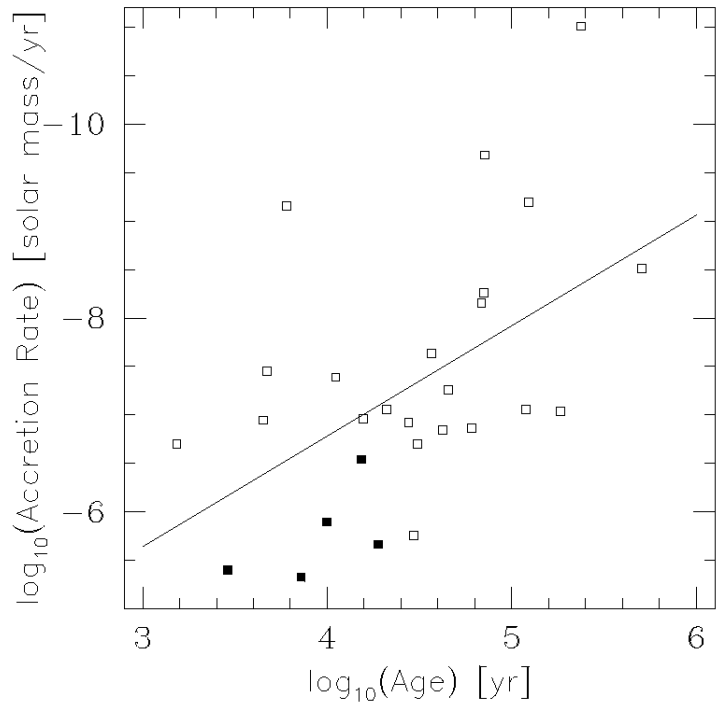}
\caption{Modeled accretion rate as a function of modeled age. The YSO ages and accretion rates are determined by the best-fit model SED. Filled in squares show sources in our sample that are catagorized as Stage I. Open squares represent objects that are Stage II YSOs. The line represents the linear fit to the data. }
\end{figure}

\clearpage

\begin{figure}
%%9
\epsscale{.6}
\plotone{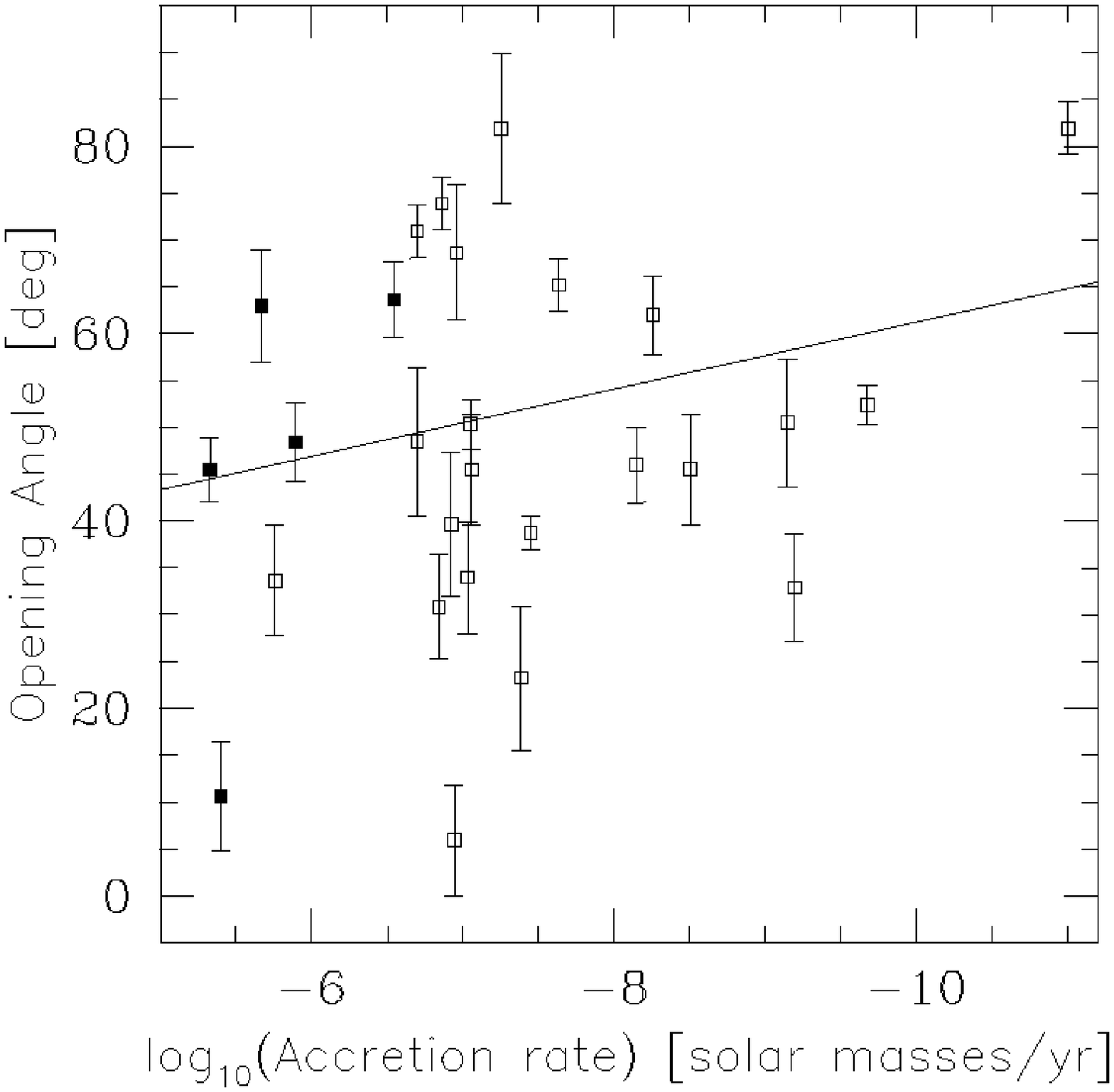}
\caption{Measured outflow opening angle as a function of modeled accretion rate. The YSO accretion rates were determined by the best-fit model SED. Opening angles are measured directly from the IRAC channel 1 images and are corrected for inclination. The uncertainties in the opening angles are from image measurement. Filled in squares show sources in our sample that are catagorized as Stage I. Open squares represent objects that are Stage II YSOs. The line represents the linear fit to the data. }
\end{figure}

\clearpage

\begin{figure}
%%10
\epsscale{1}
\plotone{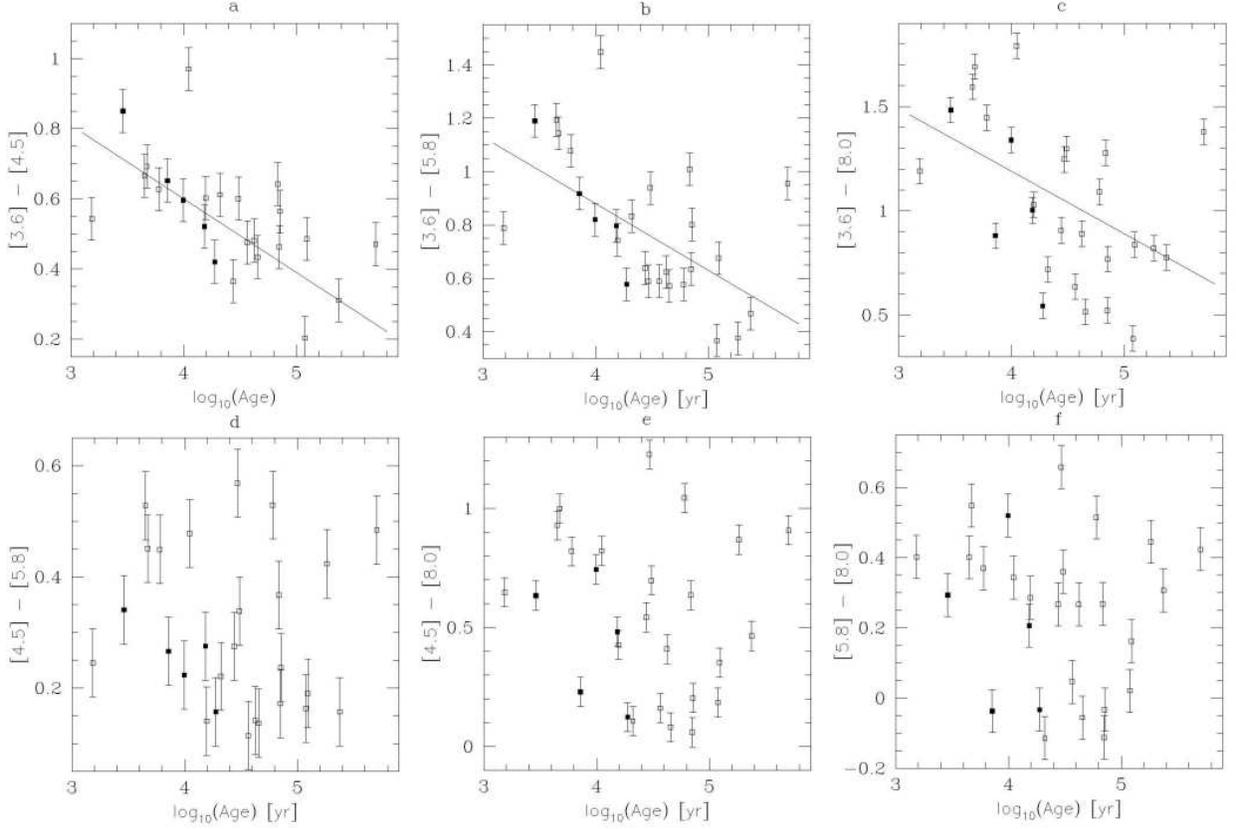}
\caption{Six different IRAC colors, [3.6]-[4.5], [3.6]-[5.8], [3.6]-[8.0], [4.5]-[5.8], [4.5]-[8.0], and [5.8]-[8.0] as a function of modeled age. The YSO colors were determined through aperture photometry at apperatures physically corresponding to 2000 AU. Filled in squares show sources in our sample that are catagorized as Stage I. Open squares represent objects that are Stage II YSOs. The line represents the linear fit to the data. }
\end{figure}

\clearpage

\begin{figure}
%%11
\epsscale{1}
\plotone{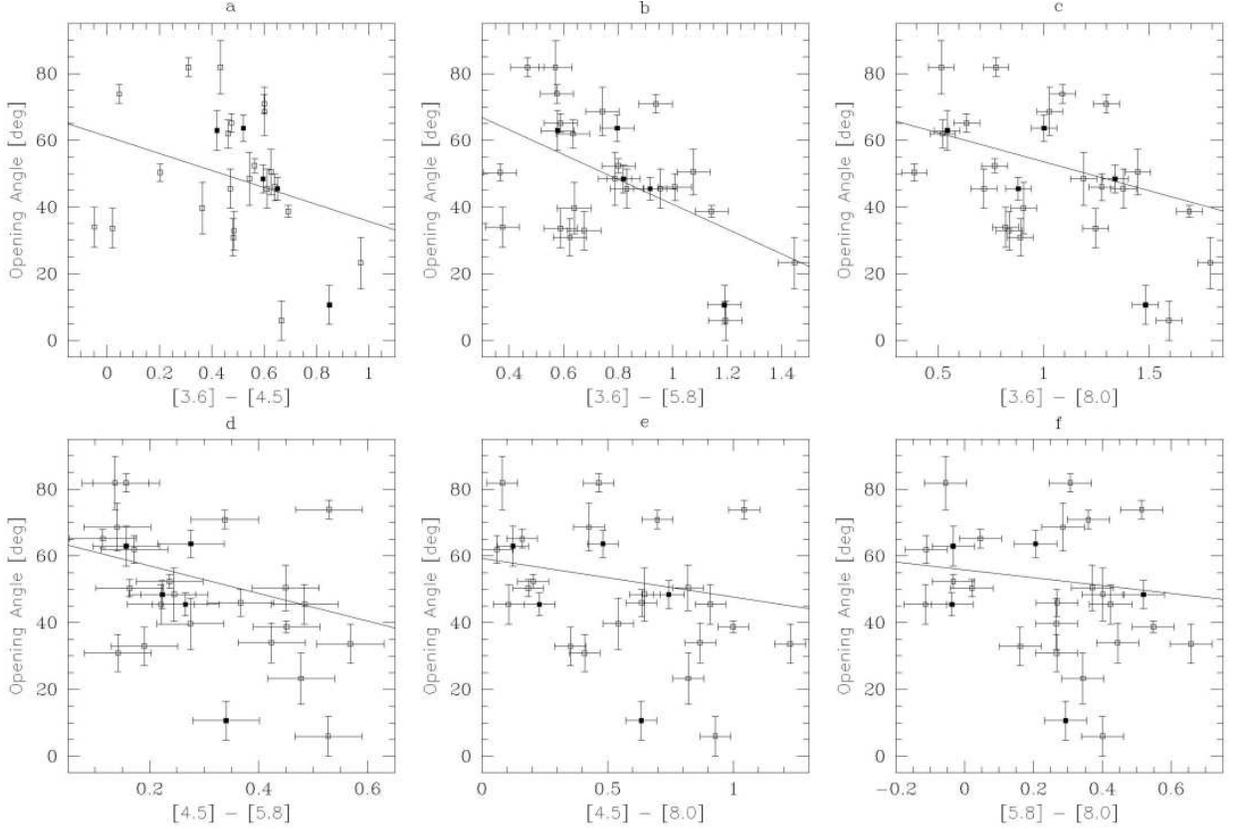}
\caption{Measured outflow opening angle as a function of six different IRAC colors, [3.6]-[4.5], [3.6]-[5.8], [3.6]-[8.0], [4.5]-[5.8], [4.5]-[8.0], and [5.8]-[8.0]. The YSO colors were determined through aperture photometry at apperatures physically corresponding to 2000 AU. Opening angles are measured directly from the IRAC channel 1 images and are corrected for inclination. The uncertainties in the opening angles are from image measurement. Filled in squares show sources in our sample that are catagorized as Stage I. Open squares show sourcese that are Stage II YSOs. The line represents the linear fit to the data. }
\end{figure}

\clearpage

\begin{deluxetable}{lccccccc}
\tabletypesize{\scriptsize}
\tablecaption{Source List}
\tablewidth{0pt}
\tablehead{
\colhead{Object} & \colhead{R.A.} & \colhead{Dec.} & \colhead{Distance (pc)} & \colhead{Distance Reference\tablenotemark{a}} & \colhead{Stage} & \colhead{Class$_{submm}$} & \colhead{Class$_{1.3 mm}$}}
\startdata
IRAS 05491+0247    & 05 51 46.1  & +02 48 30   & 460 & 1 & II& 0 & 0/I  \\

IRAS 20353+6742    & 20 35 45.9  & +67 53 02   & 440 & 2 & I & 0 & I    \\

IRAS 21169+6804    & 21 17 39.4  & +68 17 32   & 450 & 3 & II & 0 & 0/I \\

IRAS 18595$-$3712  & 19 02 58.7  & $-$37 07 35 & 170 & 4 & II & 0 & 0   \\

IRAS 03255+3103    & 03 28 37.0  & +31 13 32   & 350 & 5 & II & 0/I & I \\

IRAS 03256+3055    & 03 28 44.5  & +31 05 39   & 400 & 2 & II & 0 & 0   \\

$[$JJK2007$]$ 9        & 03 28 32.6  & +31 11 05   & 350 & 11 & II & 0 & 0  \\

IRAS 04166+2706    & 04 19 43.0  & +27 13 34   & 140 & 5 & II & 0 & 0   \\

IRAS 04368+2557	   & 04 39 53.6  & +26 03 06   & 140 & 5 & II & 0 & I   \\

IRAS 05329$-$0505  & 05 35 26.6  & $-$05 03 56 & 450 & 6 & II & 0/I & I \\

SL 05346$-$0528    & 05 34 38.1  & $-$05 27 41 & 470 & 12 & II & 0 & I  \\

IRAS 05329$-$0614  & 05 35 23.1  & $-$06 12 44 & 480 & 13 & II & 0 & 0  \\

$[$MAW2001$]$ LBS5-MM3 & 05 47 36.9  & +00 20 07   & 450 & 15 & II & I & I  \\

IRAS 23238+7401    & 23 25 45.7  & +74 17 37   & 180 & 7 & II & 0 & 0/I \\

$[$B2001b$]$ IRS 2     & 12 01 34.0  & $-$65 08 44 & 200 & 5 & I & 0 & 0/I  \\

IRAS 18148$-$0440  & 18 17 29.8  & $-$04 39 38 & 200 & 5 & II & 0 & 0/I \\

L1251B             & 22 38 46.8  & +75 11 33   & 330 & 8 & II & 0 & 0/I \\

SL 18299+0117      & 18 29 51.2  & +01 16 40   & 300 & 14 & II & 0/I & I\\

IRAS 03271+3013    & 03 30 14.9  & +30 23 48   & 320 & 9 & I & 0 & 0/I  \\

IRAS 03292+3039    & 03 32 17.6  & +30 49 50   & 320 & 9 & II & 0 & 0   \\

IRAS 03235+3004    & 03 26 37.0  & +30 15 26   & 320 & 9 & I & 0 & 0/I  \\

L1448C             & 03 25 38.8  & +30 44 05   & 300 & 5 & II & 0 & 0   \\

IRAS 03301+3057    & 03 33 16.4  & +31 07 57   & 320 & 9 & II & 0 & I   \\

HH789              & 03 33 27.5  & +31 07 36   & 320 & 9 & II & 0 & 0/I \\

IRAS 03245+3002    & 03 27 39.0  & +30 12 59   & 350 & 10 & II & 0 & 0  \\

LDN 1634 7         & 05 19 51.6  & $-$05 52 06 & 460 & 10 & II & 0/I & I\\

IRAS 05173$-$0555  & 05 19 48.9  & $-$05 52 05 & 460 & 10 & I & 0 & 0   \\

\enddata

\tablenotetext{a}{ (1) Reipurth et al. 1997, (2) Clark 1991, (3) Young et al. 2003, (4) Hamaguchi et al. 2005, (5) Motte \& Andr{\'e} 2001, (6) Andre et al. 2000, (7) Shirley et al. 2000, (8) Kun \& Prusti 1993, (9) Matthews \& Wilson 2002, (10) Persi et al. 1994, (11) Olmi et al. 2005, (12) Bally \& Reipurth 2001, (13) Zavagno et al. 1997, (14) Chavarria et al. 1988, (15) Genzel \& Stutzki 1989 }
\end{deluxetable}

\clearpage

\begin{deluxetable}{lccccc}
\tabletypesize{\scriptsize}
\tablewidth{0pt}
\tablecaption{Outflow Cavity Opening Angles}
\tablehead{\colhead{Object} & \colhead{Min Angle} & \colhead{Max Angle} & \colhead{Average Angle} & \colhead{Corrected Average Angle} & \colhead{CO Angle\tablenotemark{a}} \\
                            & \colhead{($\arcdeg$)} & \colhead{($\arcdeg$)} & \colhead{($\arcdeg$)}     & \colhead{($\arcdeg)$}               & \colhead{($\arcdeg$)}}
\startdata
IRAS 05491+0247    & 51$\pm$6 & 91$\pm$6 & 71$\pm$4 & 71$\pm$4 & \nodata \\
IRAS 20353+6742    & 63$\pm$6 & 63$\pm$6 & 63$\pm$6 & 63$\pm$6 & \nodata \\
IRAS 21169+6804    & 40$\pm$6 & 68$\pm$6 & 54$\pm$4 & 50$\pm$4 & \nodata \\
IRAS 18595$-$3712  & 46$\pm$6 & 86$\pm$6 & 66$\pm$4 & 65$\pm$4 & \nodata \\
IRAS 03255+3103    & 6$\pm$6  &  6$\pm$6 &  6$\pm$6 &  6$\pm$6 & \nodata \\
IRAS 03256+3055    & 33$\pm$8 & 33$\pm$8 & 33$\pm$8 & 31$\pm$6 & \nodata \\
$[$JJK2007$]$ 9        & 74$\pm$6 & 91$\pm$6 & 82$\pm$4 & 82$\pm$4 & \nodata \\
IRAS 04166+2706    & 34$\pm$6 & 34$\pm$6 & 34$\pm$6 & 33$\pm$6 & \nodata \\
IRAS 04368+2557    & 46$\pm$4 & 61$\pm$4 & 53$\pm$3 & 52$\pm$3 & 70\tablenotemark{b} \\
IRAS 05329$-$0505  & 63$\pm$6 & 86$\pm$6 & 74$\pm$4 & 74$\pm$4 & \nodata \\
SL 05346$-$0528    & 34$\pm$6 & 34$\pm$6 & 34$\pm$6 & 34$\pm$6 & \nodata \\
IRAS 05329$-$0614  & 74$\pm$8 & 74$\pm$8 & 74$\pm$8 & 69$\pm$7 & \nodata \\
$[$MAW2001$]$ LBS5-MM3 & 34$\pm$6 & 34$\pm$6 & 34$\pm$6 & 34$\pm$6 & \nodata \\
IRAS 23238+7401    & 74$\pm$8 & 82$\pm$8 & 78$\pm$6 & 39$\pm$3 & \nodata \\
$[$B2001b$]$ IRS 2 & 43$\pm$7 & 50$\pm$7 & 46$\pm$5 & 45$\pm$5 & \nodata \\
IRAS 18148$-$0440  & 46$\pm$6 & 46$\pm$6 & 46$\pm$6 & 45$\pm$6 & \nodata \\
L1251B             & 41$\pm$8 & 41$\pm$8 & 41$\pm$8 & 40$\pm$8 & \nodata \\
SL 18299+0117      & 46$\pm$6 & 46$\pm$6 & 46$\pm$6 & 45$\pm$6 & \nodata \\
IRAS 03271+3013    & 41$\pm$8 & 57$\pm$8 & 49$\pm$6 & 48$\pm$6 & \nodata \\
IRAS 03292+3039    & 82$\pm$8 & 82$\pm$8 & 82$\pm$8 & 82$\pm$8 & \nodata \\
IRAS 03235+3004    & 66$\pm$8 & 66$\pm$8 & 66$\pm$8 & 64$\pm$8 & \nodata \\
L1448C             & 46$\pm$4 & 46$\pm$4 & 46$\pm$4 & 46$\pm$4 & 40\tablenotemark{c}\\
IRAS 03301+3057    & 57$\pm$8 & 57$\pm$8 & 57$\pm$8 & 50$\pm$7 & \nodata \\
HH789              & 49$\pm$8 & 49$\pm$8 & 49$\pm$8 & 48$\pm$8 & \nodata \\
IRAS 03245+3002    & 24$\pm$8 & 24$\pm$8 & 24$\pm$8 & 23$\pm$8 & \nodata \\
LDN 1634 7         & 49$\pm$8 & 74$\pm$8 & 62$\pm$6 & 62$\pm$6 & \nodata \\
IRAS 05173$-$0555  & 11$\pm$6 & 11$\pm$6 & 11$\pm$6 & 11$\pm$6 & 30\tablenotemark{d}\\

\enddata
\tablenotetext{a}{Uncertainty in opening angle estimate is about 5\arcdeg--10\arcdeg}
\tablenotetext{b}{Ohashi et al.\ 1997}
\tablenotetext{c}{Bachiller et al.\ 1995}
\tablenotetext{d}{{Lee} et al.\ 2000 }
\end{deluxetable}

\newpage

\begin{deluxetable}{lccccc}
\tabletypesize{\scriptsize}
\tablewidth{0pt}
\tablecaption{IRAC Photometry}
\tablehead{\colhead{Object} & \colhead{Aperture} & \colhead{$F_{[3.6]}$} & \colhead{$F_{[4.5]}$} & \colhead{$F_{[5.8]}$} & \colhead{$F_{[8.0]}$} \\
                            & \colhead{(AU)}    & \colhead{(mJy)}     & \colhead{(mJy)}     & \colhead{(mJy)} & \colhead{(mJy)}}

\startdata
IRAS 05491+0247    & 500 & 3.98 & 12.3 & 17.0 & 17.3 \\
                   & 1000& 11.1 & 32.4 & 43.5 & 47.3 \\
                   & 2000& 21.3 & 54.3 & 75.7 & 96.6 \\
                   & 4000& 29.0 & 67.8 & 90.8 & 121.9 \\	
IRAS 20353+6742    & 500 & 1.20 & 2.04 &1.69 & 0.817 \\
                   & 1000& 2.72 & 4.66 & 3.97 & 1.94 \\
                   & 2000& 3.68 & 6.20 & 5.70 & 2.94 \\
                   & 4000& 4.69 & 7.52 & 6.70 & 3.20 \\
IRAS 21169+6804    & 500 & 18.7 & 20.8 & 17.7 & 10.3 \\
                   & 1000& 52.0 & 56.2 & 50.1 & 27.9 \\
                   & 2000& 87.0 & 88.9 & 82.8 & 48.5 \\
                   & 4000& 121.0 & 118.0 & 106.8 & 62.0 \\
IRAS 18595$-$3712  & 500 & 7.58 & 17.5 & 16.2 & 9.77 \\
                   & 1000& 12.1 & 25.7 & 23.7 & 14.8 \\
                   & 2000& 18.1 & 34.6 & 28.8 & 17.9 \\
                   & 4000& 26.2 & 45.6 & 27.0 & 17.3 \\
IRAS 03255+3103    & 500 & 15.6 & 48.8 & 109.7 & 19.9 \\
                   & 1000& 32.6 & 101.3 & 219.6 & 131.2 \\
                   & 2000& 45.6 & 135.2 & 291.7 & 409.2 \\
                   & 4000& 55.8 & 157.4 & 328.6 & 561.5 \\
IRAS 03256+3055    & 500 & 0.647 & 1.55 & 1.11 & 1.40 \\
                   & 1000& 1.40 & 2.94 & 2.50 & 2.38 \\
                   & 2000& 1.95 & 3.78 & 3.35 & 3.46 \\
                   & 4000& 1.89 & 3.40 & 3.07 & 3.04 \\
$[$JJK2007$]$ 9    & 500 & 0.734 & 1.17 & 1.16 & 1.36 \\
                   & 1000& 1.65 & 2.43 & 2.43 & 2.68 \\
                   & 2000& 2.69 & 3.53 & 3.24 & 3.67 \\
                   & 4000& 3.75 & 4.64 & 3.53 & 3.85 \\
IRAS 04166+2706    & 500 & 1.21 & 4.46 & 6.07 & 6.21 \\
                   & 1000& 4.06 & 9.50 & 11.1 & 10.3 \\
                   & 2000& 7.71 & 15.1 & 15.0 & 12.1 \\
                   & 4000& 10.6 & 18.6 & 16.2 & 13.7 \\
IRAS 04368+2557    & 500 & 2.39 & 9.94 & 17.1 & 11.5 \\
                   & 1000& 7.08 & 21.9 & 32.3 & 20.7 \\
                   & 2000& 24.2 & 56.7 & 62.6 & 32.4 \\
                   & 4000& 63.9 & 122.8& 112.7& 52.0 \\
IRAS 05329$-$0505  & 500 & 31.0 & 5.10 & 3.79 & 5.42 \\
                   & 1000 & 86.7  & 33.7 & 42.4 & 67.6 \\
                   & 2000 & 138.5  & 98.7 & 213.7 & 389.2 \\
                   & 4000 & 186.0  & 159.5 & 346.0 & 609.7 \\
SL 05346$-$0528    & 500 & 1.82  & 1.19 & 1.31 & 2.26 \\
                   & 1000 & 4.27  & 2.97 & 3.88 & 7.90 \\
                   & 2000 & 6.70  & 4.49 & 10.6 & 27.0 \\
                   & 4000 & 10.7  & 6.65 & 28.5 & 79.8 \\
IRAS 05329$-$0614  & 500 & 0.396  & 1.26 & 0.915 & 0.656 \\
                   & 1000 & 1.02  & 2.94 & 2.15 & 1.69 \\
                   & 2000 & 1.77 & 4.55 & 4.02 & 4.33 \\
                   & 4000 & 3.16  & 6.31 & 7.39 & 11.9 \\
$[$MAW2001$]$ LBS5-MM3 & 500 & 36.5  & 5.09 & 11.6 & 10.8 \\
                   & 1000 & 112.0 & 35.8 & 68.8 & 74.9 \\
                   & 2000 & 208.8 & 119.6 & 202.9 & 315.1 \\
                   & 4000 & 290.0  & 191.1 & 283.8 & 461.5 \\
IRAS 23238+7401    & 500 & 4.10 & 14.1 & 26.4 & 47.1 \\
                   & 1000 & 5.59  & 18.4 & 34.6 & 70.1 \\
                   & 2000 & 6.93 & 21.8 & 39.5 & 77.8 \\
                   & 4000 & 7.93  & 24.4 & 40.7 & 69.6 \\
$[$B2001b$]$ IRS 2 & 500 & 2.35 & 6.54 & 7.84 & 4.62 \\
                   & 1000 & 4.10  & 11.7 & 14.8 & 8.68 \\
                   & 2000 & 6.08  & 17.4 & 20.6 & 10.5 \\
                   & 4000 & \nodata & \nodata & \nodata & \nodata \\
IRAS 18148$-$0440  & 500 & 16.2  & 65.3 & 67.7 & 28.9 \\
                   & 1000 & 32.9  & 116.9 & 129.6 & 56.4 \\
                   & 2000 & 74.8  & 195.6 & 208.3 & 89.3 \\
                   & 4000 & 129.8  & 286.8 & 270.5 & 104.1 \\
L1251B             & 500 & 3.29 & 5.31 & 6.00 & 6.46 \\
                   & 1000 & 6.11  & 9.71 & 11.6 & 11.7 \\
                   & 2000 & 9.34  & 13.8 & 16.6 & 17.1 \\
                   & 4000 & 15.9 & 23.2 & 26.6 & 22.3 \\
SL 18299+0117      & 500 & 16.0  & 34.4 & 62.6 & 72.0 \\
                   & 1000 & 28.2 & 58.1 & 113.4 & 141.2 \\
                   & 2000 & 36.2 & 68.6 & 133.9 & 198.3 \\
                   & 4000 & 42.0 & 78.4 & 138.1 & 189.7 \\
IRAS 03271+3013    & 500 & 3.95 & 10.0 & 9.92 & 17.1 \\
                   & 1000 & 6.46 & 16.9 & 17.8 & 30.3 \\
                   & 2000 & 8.07 & 20.3 & 21.8 & 40.2 \\
                   & 4000 & 9.71 & 23.9 & 22.8 & 38.6 \\
IRAS 03292+3039    & 500 & 0.626 & 1.17 & 0.971 & 0.525 \\
                   & 1000 & 1.56 & 2.87 & 2.53 & 1.36 \\
                   & 2000 & 2.53 & 4.39 & 3.85 & 1.89 \\
                   & 4000 & 3.15 & 5.60 & 3.93 & 0.630 \\
IRAS 03235+3004    & 500 & 2.01 & 4.74 & 5.49 & 4.92 \\
                   & 1000 & 3.52 & 8.33 & 10.0 & 8.65 \\
                   & 2000 & 4.81 & 10.2 & 12.3 & 11.0 \\
                   & 4000 & \nodata & \nodata & \nodata & \nodata \\
L1448C             & 500 & 2.33 & 7.83 & 10.2 & 9.53 \\
                   & 1000 & 5.16 & 16.0 & 22.2 & 20.2 \\
                   & 2000 & 8.93 & 25.0 & 37.3 & 38.6 \\
                   & 4000 & \nodata & \nodata & \nodata & \nodata \\
IRAS 03301+3057    & 500 & 8.65 & 24.8 & 40.7 & 46.5 \\
                   & 1000 & 14.6 & 41.6 & 73.4 & 84.7 \\
                   & 2000 & 17.7 & 47.9 & 86.4 & 113.1 \\
                   & 4000 & 20.6 & 54.8 & 91.7 & 108.9 \\
HH789              & 500 & 1.42 & 3.37 & 3.31 & 4.99 \\
                   & 1000 & 2.52 & 5.98 & 6.45 & 9.07 \\
                   & 2000 & 3.26 & 7.29 & 8.20 & 11.5 \\
                   & 4000 & 4.00 & 8.63 & 8.82 & 10.2 \\
IRAS 03245+3002    & 500 & 0.331 & 3.40 & 7.09 & 8.49 \\
                   & 1000 & 0.797 & 7.14 & 14.9 & 16.7 \\
                   & 2000 & 1.88 & 11.2 & 21.6 & 26.5 \\
                   & 4000 & \nodata & \nodata & \nodata & \nodata \\
LDN 1634 7         & 500 & 4.26 & 8.75 & 6.73 & 2.45 \\
                   & 1000 & 10.7 & 20.2 & 17.4 & 6.72 \\
                   & 2000 & 15.9 & 29.6 & 28.2 & 12.1 \\
                   & 4000 & 21.1 & 36.9 & 35.5 & 16.2 \\
IRAS 05173$-$0555  & 500 & 0.568 & 2.94 & 3.86 & 4.51 \\
                   & 1000 & 1.46 & 7.47 & 9.84 & 10.8 \\
                   & 2000 & 2.50 & 11.3 & 15.9 & 17.4 \\
                   & 4000 & 4.41 & 15.5 & 21.2 & 21.5 \\
\enddata

\end{deluxetable}

\clearpage

\begin{deluxetable}{lcccccccc}
\tabletypesize{\scriptsize}
\tablewidth{0pt}
\tablecaption{MIPS \& IRAS Photometry}
\tablehead{\colhead{Object} & \colhead{24 $\mu$m Aperture} & \colhead{$F_{[24]}$} & \colhead{70 $\mu$m Aperture}  & \colhead{$F_{[70]}$} & \colhead{$F_{[12]}$} & \colhead{$F_{[25]}$} & \colhead{$F_{[60]}$} & \colhead{$F_{[100]}$} \\
                            & \colhead{(arcsec)}          & \colhead{(Jy)}     & \colhead{(arcsec)}           & \colhead{(Jy)}     & \colhead{(Jy)}  & \colhead{(Jy)}    & \colhead{(Jy)}    & \colhead{(Jy)}}
\startdata
IRAS 05491+0247    & 19.6 & 5.49 & 48.0 & 35.5 & 0.28 & 6.74 & 43.86 & 72.14 \\
IRAS 20353+6742    & \nodata & \nodata & \nodata & \nodata & 0.25L & 0.34 & 4.09 & 7.14 \\
IRAS 21169+6804    & \nodata & \nodata & \nodata & \nodata & 0.25L & 0.68 & 11.75 & 33.53 \\
IRAS 18595$-$3712  & 19.6 & 2.31 & 60.0 & 29.1 & 0.27L & 3.69 & 38.91 & 95.24L \\
IRAS 03255+3103    & \nodata & \nodata & \nodata & \nodata & 0.56 & 16.45 & 117.30L & 531.80L \\
IRAS 03256+3055    & \nodata & \nodata & \nodata & \nodata & 0.45L & 0.27L & 1.43 & 12.75L \\
$[$JJK2007$]$ 9    & 17.15 & 0.056 & \nodata & \nodata & \nodata & \nodata & \nodata & \nodata \\
IRAS 04166+2706    & 17.15 & 0.462 & \nodata & \nodata & 0.25L & 0.57 & 5.01 & 17.46L \\
IRAS 04368+2557    & 19.6 & 0.536 & \nodata & \nodata & 0.25L & 0.74 & 17.77 & 73.26 \\
IRAS 05329$-$0505  & 17.15 & 9.54 & 28.0 & 77.9 & 4.48 & 32.01 & 84.93L & 27.28L \\
SL 05346$-$0528    & \nodata & \nodata & \nodata & \nodata & \nodata & \nodata & \nodata & \nodata \\
IRAS 05329$-$0614  & \nodata & \nodata & \nodata & \nodata & 0.60L & 0.60L & 5.04 & 49.12L \\
$[$MAW2001$]$ LBS5-MM3 & 19.6 & 6.60 & \nodata & \nodata & \nodata & \nodata & \nodata & \nodata \\
IRAS 23238+7401    & 17.15 & 1.58 & 48.0 & 10.6 & 0.25L & 0.78 & 9.60 & 15.20 \\
$[$B2001b$]$ IRS 2 & \nodata & \nodata & \nodata & \nodata & \nodata & \nodata & \nodata & \nodata \\
IRAS 18148$-$0440  & \nodata & \nodata & \nodata & \nodata & 0.25L & 6.91 & 89.05 & 165.50 \\
L1251B             & \nodata & \nodata & \nodata & \nodata & \nodata & \nodata & \nodata & \nodata \\
SL 18299+0117      & 14.7 & 1.05 & 16.0 & 5.89 & \nodata & \nodata & \nodata & \nodata \\
IRAS 03271+3013    & 17.15 & 1.80 & \nodata & \nodata & 0.25L & 1.71 & 7.53 & 8.19 \\
IRAS 03292+3039    & \nodata & \nodata & \nodata & \nodata & 0.52L & 0.31L & 1.59 & 9.09 \\
IRAS 03235+3004    & 17.15 & 0.375 & 12.0 & 2.68 & 0.25L & 0.54 & 4.18 & 6.97 \\
L1448C             & 22.05 & 2.36 & 48.0 & 38.9 & \nodata & \nodata & \nodata & \nodata \\
IRAS 03301+3057    & 17.15 & 1.78 & 24.0 & 5.31 & 0.25L & 1.39 & 6.18L & 35.47 \\
HH789              & 17.15 & 1.55 & 20.0 & 3.80 & \nodata & \nodata & \nodata & \nodata \\
IRAS 03245+3002    & 19.6 & 1.88 & 40.0 & 29.0 & 0.25L & 3.40 & 47.13 & 93.57 \\
LDN 1634 7         & \nodata & \nodata & \nodata & \nodata & \nodata & \nodata & \nodata & \nodata \\
IRAS 05173$-$0555  & \nodata & \nodata & \nodata & \nodata & 0.25L & 3.02 & 27.15 & 61.34 \\

\enddata
\tablecomments{IRAS photometry marked with an `L' is an upper limit. Flux measurements for 12 $\mu$m, 25 $\mu$m, 60 $\mu$m, and 100 $\mu$m have apertures of 60'', 60'', 120'', and 120'', respectively.}

\end{deluxetable}

\clearpage

\begin{deluxetable}{ccccc}
\tabletypesize{\scriptsize}
\tablewidth{0pt}
\tablecaption{Correlation and Rank Coefficients}
\tablehead{\colhead{Figure} & \colhead{r\tablenotemark{a}} & \colhead{P$_{>r}$\tablenotemark{b}} & \colhead{$\tau$\tablenotemark{c}}  & \colhead{P$_{\tau}$\tablenotemark{d}}  \\
                            &             & \colhead{(\%)}     & \colhead{}      & \colhead{(\%)} }
\startdata
4 &	0.34 & 		4.0 &		0.18 & 		19 			\\
5 & 	0.19 &		17  & 		0.06 & 		63 			\\
6 & 	-0.78 & 	$<10^{-5}$ & 	-0.59 & 	$1.6 \times 10^{-3}$ 	\\
7 & 	-0.29 & 	7.4 & 		-0.30 & 	2.9 			\\
8 &	-0.52 &		0.21 & 		-0.41 &		0.25			\\
9 &	-0.27 &		9.0 & 		-0.09 & 	49			\\
10a & 	-0.56 & 	0.09 & 		-0.47 & 	0.06			\\
10b & 	-0.57 & 	0.08 & 		-0.43 & 	0.16 			\\
10c &	-0.48 & 	0.47 & 		-0.36 & 	0.81			\\
10d &	-0.14 & 	25 &		-0.14 &		31			\\
10e & 	-0.16 & 	21 & 		-0.16 & 	23			\\
10f & 	-0.16 & 	21 & 		-0.15 & 	27			\\
11a & 	-0.30 &		6.1 & 		-0.26 & 	5.2			\\
11b & 	-0.51 & 	0.28 & 		-0.37 & 	0.63			\\
11c &	-0.52 & 	0.22 & 		-0.32 & 	2.1 			\\
11d & 	-0.46 & 	0.80 & 		-0.22 & 	11			\\
11e & 	-0.38 & 	2.3 & 		-0.26 & 	5.2 			\\
11f & 	-0.29 &		6.9 & 		-0.17 & 	20			\\

\enddata

\tablenotetext{a}{ Pearson product-moment correlation coefficient }
\tablenotetext{b}{ Probability that 27 points selected from a random distribution of data have a Pearson product-moment correlation coefficient of equal sign and greater than or equal magnitude than that determined by the data, r }
\tablenotetext{c}{ Kendall's Tau rank correlation coefficient }
\tablenotetext{d}{ Probability of no correlation as determined by Kendall's Tau }

\end{deluxetable}

\clearpage

\end{document}